\documentclass[epj]{webofc}
\usepackage[utf8]{inputenc}
\usepackage[varg]{txfonts}   
\usepackage{booktabs}
\usepackage{epsfig}
\usepackage{bm}
\usepackage{xcolor}
\definecolor{darkred}{rgb}{0.4,0.0,0.0}
\definecolor{darkgreen}{rgb}{0.0,0.4,0.0}
\definecolor{darkblue}{rgb}{0.0,0.0,0.4}
\usepackage[bookmarks,linktocpage,colorlinks,
    linkcolor = darkred,
    urlcolor  = darkblue,
    citecolor = darkgreen]{hyperref}

\usepackage{subfigure}
\wocname{EPJ Web of Conferences}
\woctitle{Lattice2017}
\begin{document}
%
\selectlanguage{english}
\title{%
Complex Langevin Simulations of QCD at Finite Density -- Progress Report
}
\author{%
\firstname{D.~K.} \lastname{Sinclair}\inst{1}\fnsep\thanks{Speaker, 
\email{dks@hep.anl.gov} This research was supported in part by US Department 
of Energy contract DE-AC02-06CH11357} \and
\firstname{J.~B.} \lastname{Kogut}\inst{2,3} 
}
\institute{%
HEP Division, Argonne National Laboratory, 9700 South Cass Avenue, Argonne, 
Illinois 60439, USA
\and
Department of Energy, Division of High Energy Physics, Washington, DC 20585,
USA
\and
Department of Physics -- TQHN, University of Maryland, 82 Regents Drive, 
College Park, MD 20742, USA
}
\abstract{%
We simulate lattice QCD at finite quark-number chemical potential to study 
nuclear matter, using the complex Langevin equation (CLE). The CLE is
used because the fermion determinant is complex so that standard methods 
relying on importance sampling fail. Adaptive methods and gauge-cooling are 
used to prevent runaway solutions. Even then, the CLE is not guaranteed to give
correct results. We are therefore performing extensive testing to determine
under what, if any, conditions we can achieve reliable results. Our earlier
simulations at $\beta=6/g^2=5.6$, $m=0.025$ on a $12^4$ lattice reproduced the 
expected phase structure but failed in the details. Our current simulations
at $\beta=5.7$ on a $16^4$ lattice fail in similar ways while showing some
improvement. We are therefore moving to even weaker couplings to see if the
CLE might produce the correct results in the continuum (weak-coupling) limit,
or, if it still fails, whether it might reproduce the results of the 
phase-quenched theory. We also discuss action (and other dynamics) 
modifications which might improve the performance of the CLE.
}
\maketitle
\section{Introduction}\label{intro}

We study the physics of nuclear matter, the constituent of neutron stars and
the interiors of heavy nuclei. In addition, hot nuclear matter is produced in
relativistic heavy-ion colliders, and was present in the early universe.
Ignoring the important effects of the electromagnetic field, nuclear matter is
QCD at finite quark-/baryon-number density.

Lattice QCD at a finite quark-number chemical potential $\mu$ has a complex
fermion determinant. Hence standard simulation methods based on importance
sampling fail. The Langevin approach does not rely on importance sampling and
can be extended to complex actions
\cite{Parisi:1984cs,Klauder:1983nn,Klauder:1983zm,Klauder:1983sp}. 
For lattice QCD, this requires analytic
continuation of the gauge fields from $SU(3)$ to $SL(3,C)$. However, complex
Langevin (CLE) simulations cannot be guaranteed to work unless the
trajectories are restricted to a finite domain and the drift (force) term is
holomorphic in the fields. Detailed discussion of these requirements as they
pertain to QCD at finite $\mu$ are described in
\cite{Aarts:2009uq,Aarts:2011ax,Nagata:2015uga,Nishimura:2015pba,Nagata:2016vkn,
Aarts:2017vrv,Seiler:2017wvd}.
In lattice QCD at weak coupling, potential runaway solutions are controlled by
adaptively readjusting the updating `time' increment and implementing gauge
cooling -- choosing a gauge which minimizes the average unitarity norm, which
is a measure of the distance of the gauge fields from the $SU(3)$ manifold
\cite{Seiler:2012wz}.

For lattice QCD at finite $\mu$, the gauge-cooled CLE trajectories do appear
to be confined to compact domains. However, zeros of the fermion determinant
produce poles in the drift term, so that it is meromorphic not holomorphic in
the fields. Hence observables cannot be guaranteed to converge to the correct
limits. Simulations of heavy-quark lattice QCD where the hopping parameter
expansion is used, have been performed using the CLE
\cite{Aarts:2008rr,Aarts:2013uxa,Aarts:2014bwa,Aarts:2016qrv,Langelage:2014vpa}.
More limited simulations have been performed for lighter quark masses
\cite{Sexty:2013ica,Aarts:2014bwa,Fodor:2015doa,Nagata:2016mmh}. 
We have been testing the CLE for lattice QCD at finite $\mu$ at zero 
temperature, over a range of $\mu$ values from $0$ to saturation.

Our earlier simulations at $\beta=6/g^2=5.6$, $m=0.025$ mainly on $12^4$
lattices show some of the properties expected of finite density QCD, but fail
in the details \cite{Sinclair:2015kva,Sinclair:2016nbg}. 
We are now extending these to weaker coupling, $\beta=5.7$, on
a $16^4$ lattice, with $m=0.025$. While this improves the agreement with known
(RHMC) observables at $\mu=0$, the behaviour for $\mu >0$ through the
transition region shows similar deficiencies to those at $\beta=5.6$ albeit
with some improvement. For $\mu$ values above the transition region observables
from our $\beta=5.7$ CLE simulations agree with expectations whereas
$\beta=5.6$ exhibits small discrepancies.

One important improvement over $\beta=5.6$ is that the unitarity norm is
significantly smaller for $\beta=5.7$ than for $\beta=5.6$. Since the
experience of others has identified keeping this norm low as a a way of
improving CLE results, this encourages one to try even weaker couplings
requiring even larger lattices.

Recent results from random matrix theory indicate that, when the complex 
Langevin fails, it produces results consistent with the phase-quenched theory
\cite{Bloch:2016jwt}.   
Preliminary indications from our simulations are that this is not true for
lattice QCD at finite $\mu$ for these couplings. It is still an open question
as to whether the CLE produces the correct results, the phase-quenched results
or neither in the weak-coupling limit. Other random matrix CLE simulations
seem more optimistic \cite{Nagata:2016alq}. Here it is emphasized that
gauge-cooling is essential for obtaining the correct results.

\section{Complex Langevin Equation for finite density Lattice QCD}

If $S(U)$ is the gauge action obtained from integrating out the quark fields, 
the Langevin equation for the evolution of the gauge fields $U$ in Langevin 
time $t$ is:
\begin{equation}
-i \left(\frac{d}{dt}U_l\right)U_l^{-1} = -i \frac{\delta}{\delta U_l}S(U)
+\eta_l
\end{equation}
where $l$ labels the links of the lattice, and 
$\eta_l=\eta^a_l\lambda^a$. Here $\lambda_a$ are the Gell-Mann 
matrices for $SU(3)$. $\eta^a_l(t)$ are Gaussian-distributed random 
numbers normalized so that:
\begin{equation}
\langle\eta^a_l(t)\eta^b_{l'}(t')\rangle=\delta^{ab}\delta_{ll'}\delta(t-t')
\end{equation}

The complex-Langevin equation has the same form except that the $U$s are now
in $SL(3,C)$. $S$, now $S(U,\mu)$ is 
\begin{equation}
S(U,\mu) = \beta\sum_{_\Box} \left\{1-\frac{1}{6}{\rm Tr}[UUUU+(UUUU)^{-1}]
\right\} - \frac{N_f}{4}{\rm Tr}\{\ln[M(U,\mu)]\}
\end{equation}
where $M(U,\mu)$ is the staggered Dirac operator. Note: backward links
are represented by $U^{-1}$ not $U^\dag$. Note also that we have 
chosen to keep the noise-vector $\eta$ real. $\eta$ is gauge-covariant under
$SU(3)$, but not under $SL(3,C)$. This means that gauge-cooling is non-trivial.
Reference~\cite{Nagata:2015uga} indicates why this is not expected to change
the physics. After taking $-i\delta S(U,\mu)/\delta U_l$, the cyclic
properties of the trace are used to rearrange the fermion term so that it
remains real for $\mu=0$ even after replacing the trace by a stochastic
estimator, at least for infinite precision.

To simulate the time evolution of the gauge fields we use the partial 
second-order formalism of Fukugita, Oyanagi and Ukawa.
\cite{Ukawa:1985hr,Fukugita:1986tg,Fukugita:1988qs} 

After each update, we gauge-fix iteratively to a gauge which minimizes the 
unitarity norm:
\begin{equation}
F(U) = \frac{1}{4V}\sum_l{\rm Tr}\left[U_l^\dag U_l + (U_l^\dag U_l)^{-1} 
     - 2\right] \ge 0,
\end{equation}
where $V$ is the space-time volume of the lattice. This is referred to as gauge 
cooling \cite{Seiler:2012wz}.

\section{Zero Temperature Simulations on a $16^4$ lattice at 
$\beta=5.7$}

Since our earlier CLE simulations of lattice QCD at $\beta=5.6$, $m=0.025$ on
a $12^4$ lattice correctly reproduced some aspects of the expected phase
structure, but failed in the details, we are repeating such simulations at
weaker coupling (while continuing our $\beta=5.6$ runs). We are running
simulations at $\beta=5.7$, $m=0.025$. In order to be sure that the $\mu=0$
theory is confined, we have increased our lattice size to $16^4$. Important
$\mu$ values are $m_\pi/2 \approx 0.194$ and $m_N/3 \approx 0.28$. These
were obtained from results published by the Columbia group
\cite{Brown:1991qw,Schaffer:1992rq}.
We are running 2 or 3 million updates at each $\mu$, from $\mu=0$ up to
saturation. Our input $dt=0.01$, and we perform 5 gauge-cooling steps after
each update. After adaptive rescaling of $dt$ this gives us from
$\approx 80$ to over $1000$ equilibrated time-units per $\mu$.

A recent paper by Bloch {\it et al.} \cite{Bloch:2016jwt}
has demonstrated that when CLE simulations
of a random matrix theory at finite $\mu$, which is a model for QCD at
finite $\mu$, converges to the wrong limit, it produces the results of the
phase-quenched theory. We therefore run RHMC simulations of phase-quenched
lattice QCD on the same size lattice at the same $\beta$ over the same range
of $\mu$ values. (For a description of our methods and the physics of 
phase-quenched QCD see for example \cite{Kogut:2002zg,Sinclair:2006zm}.)
These results are plotted on the same graphs as the CLE
data, for comparison. Note that for these phase-quenched simulations we need
to introduce a small symmetry breaking parameter $\lambda$. We will eventually
need to run at more than one choice of this parameter and extrapolate to 
$\lambda=0$.

The physics one expects is that all observables should remain at their $\mu=0$
values up to the transition. For the full theory, this transition should occur
at $\mu$ just below $m_N/3$, while for the phase-quenched theory it should
occur at $\mu \approx m_\pi/2$. For the the phase-quenched theory, the
transition should be second order, while for the full theory it is expected to
be first order. Beyond this  phase transition the chiral condensate
($\langle\bar{\psi}\psi\rangle$) should fall rapidly approaching zero at
saturation, if not before. The quark number density should remain zero up to
the transition. Above this it should rise, approaching $3$ at saturation. The
plaquette should remain constant up to the transition. Above this it should
increase towards its saturation value. It is expected that, when $\mu$ is
sufficiently large, the phase of the fermion determinant will cease to be
important, and the observables of the full and phase-quenched theories should
be identical.

\begin{figure}[htb]
\parbox{2.75in}{
\epsfxsize=2.75in
\epsffile{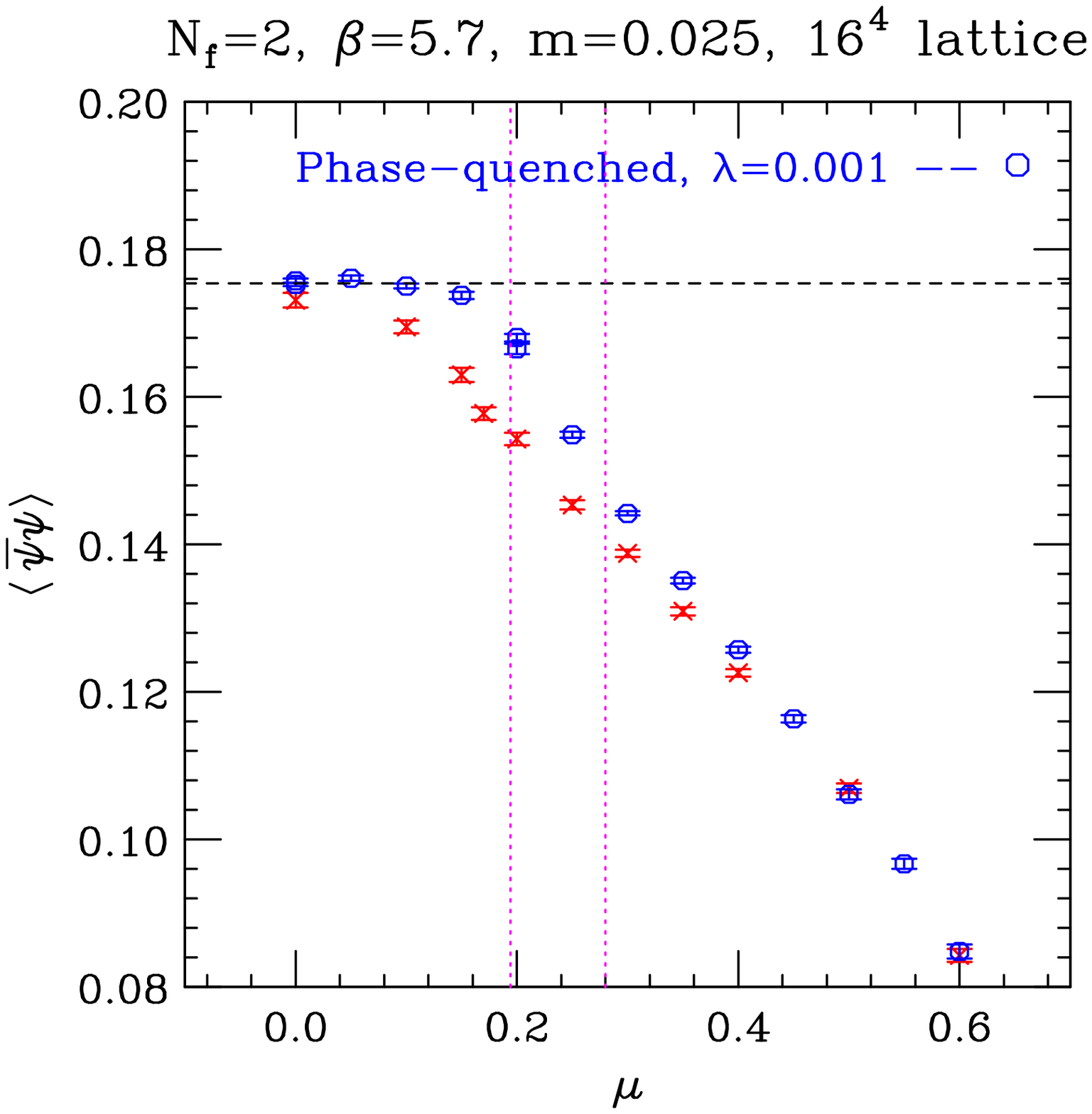}
\caption{Chiral condensates as functions of $\mu$ for $\beta=5.7$ on a $16^4$
lattice. Red crosses, full(CLE) QCD; blue circles, phase-quenched QCD. Vertical
magenta dotted lines are at $m_\pi/2$ and $m_N/3$.}
\label{fig:pbp57}
}
\parbox{0.2in}{}
\parbox{2.75in}{
\epsfxsize=2.75in   
\epsffile{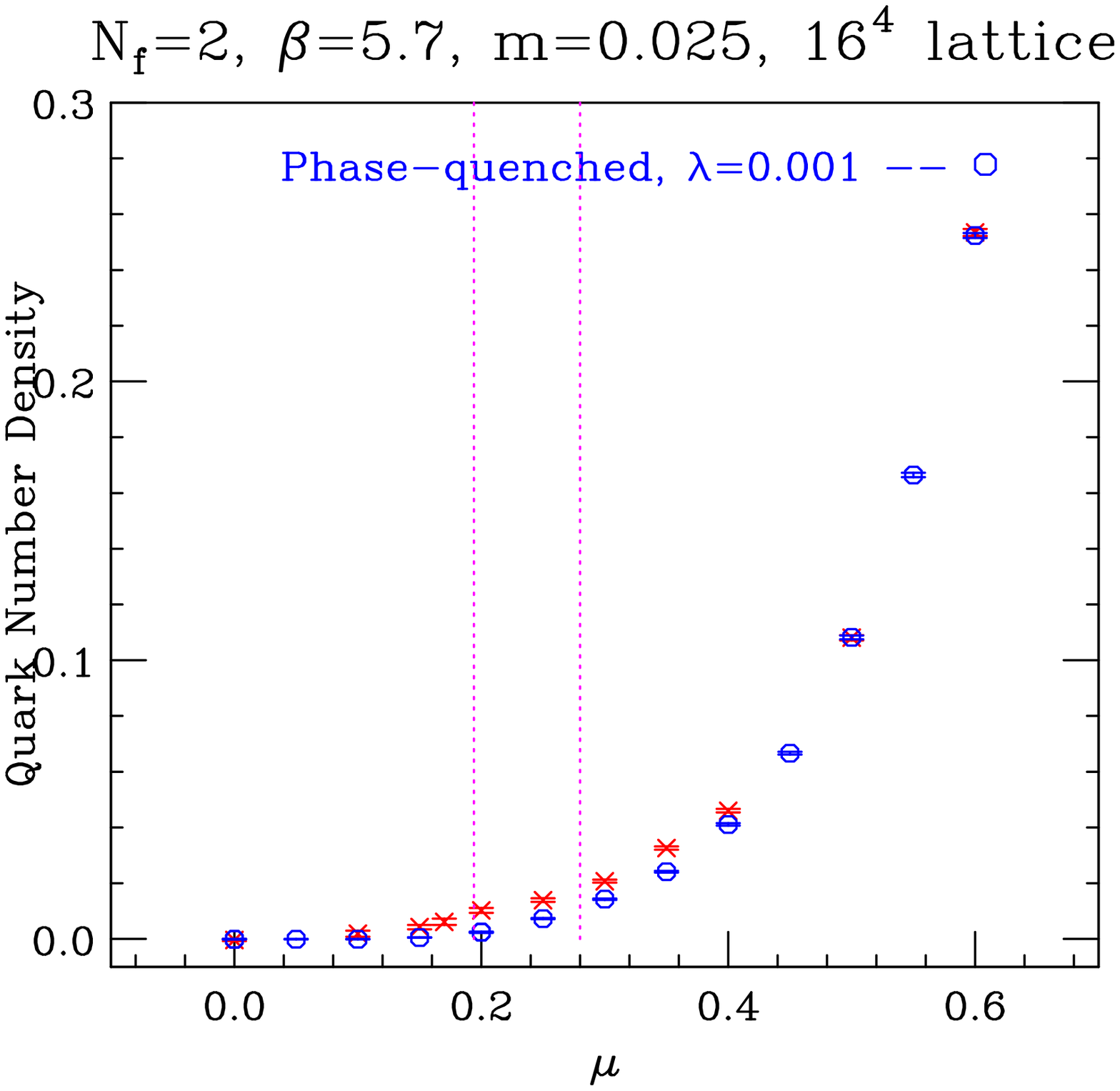}
\caption{Quark-number densities as functions of $\mu$ for $\beta=5.7$ on a 
$16^4$ lattice. Red crosses, full(CLE) QCD; blue circles, phase-quenched QCD.
Vertical magenta dotted lines are at $m_\pi/2$ and $m_N/3$.}
\label{fig:qnd57}
}
\end{figure}

Figure~\ref{fig:pbp57} shows the chiral condensate,
$\langle\bar{\psi}\psi\rangle$, from our $\beta=5.7$ CLE simulations of the
full theory and the phase-quenched results. We note that there is good
agreement at $\mu=0$, which was not true at $\beta=5.6$. However the
condensate from our CLE simulations starts to fall (almost) immediately $\mu$
is increased from zero disagreeing with both the expected behaviour and the
phase-quenched simulations. However, there is some indication that it falls
more slowly than at $\beta=5.6$. The full (CLE) and phase-quenched results
converge as $\mu$ is increased until by $\mu=0.5$ they are statistically
indistinguishable. Similar convergence is observed for $\beta=5.6$. Such large
$\mu$ behaviour agrees with expectations.

Figure~\ref{fig:qnd57} shows the quark-number density from these simulations
both of the full theory and of the phase-quenched approximation. While there is
good agreement between the full (CLE) and phase quenched results for very small
$\mu$, where both are consistent with zero, the 2 diverge for $0.1 < \mu < 0.5$.
The difference is small numerically, only because this quantity is small in 
this region. For $\mu \ge 0.5$ the results for the 2 theories are statistically
indistinguishable. This contrasts with the $\beta=5.6$ case where a small but
significant difference persists up to saturation 
\footnote{In the graph of quark-number densities at $\beta=5.6$ presented at
the conference the wrong observable was plotted for the phase-quenched theory.
The correct observable is much closer to the full (CLE) observable.}.

\begin{figure}[htb]
\parbox{2.75in}{
\epsfxsize=2.75in
\epsffile{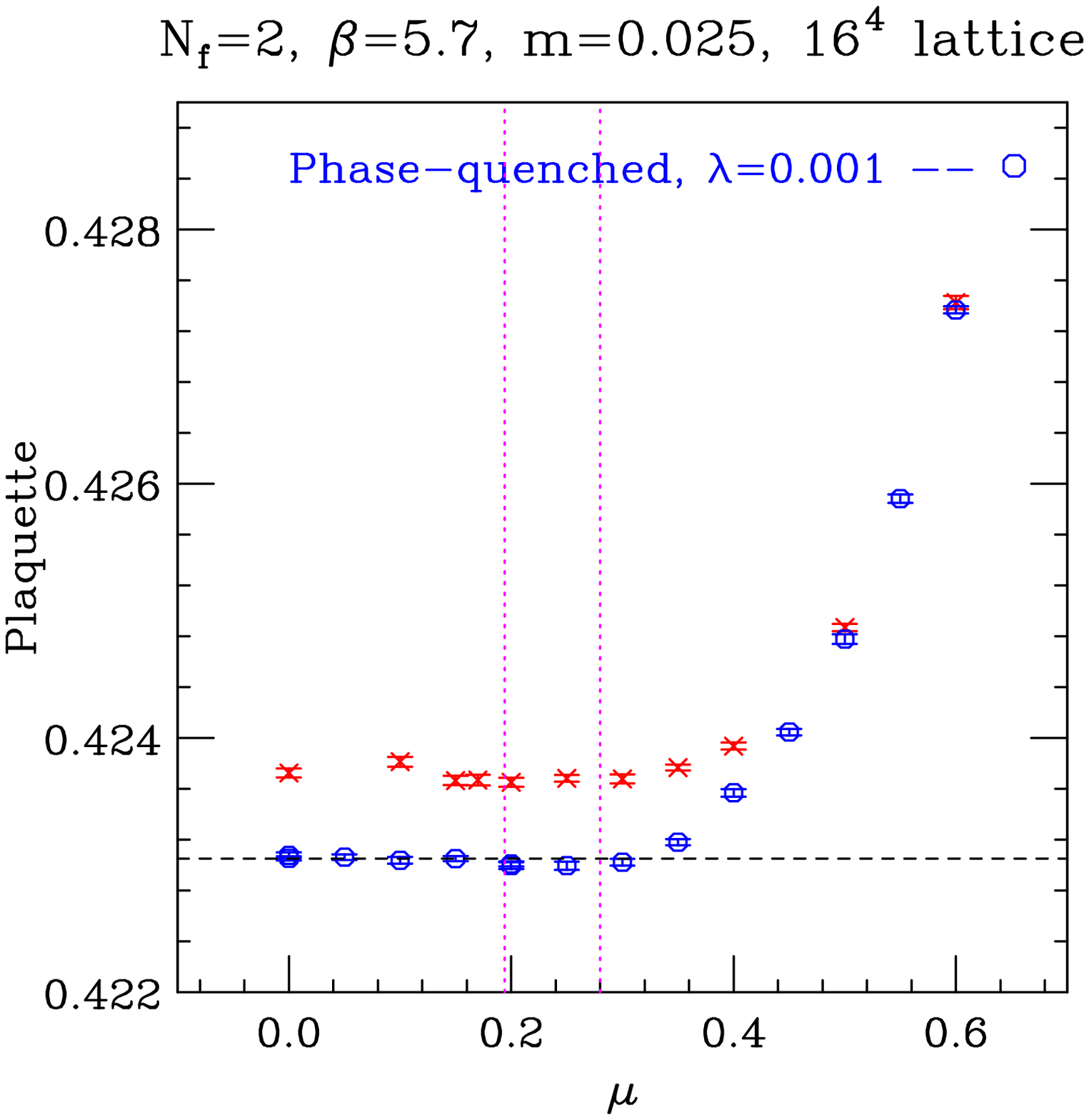}
\caption{Plaquettes as functions of $\mu$ for $\beta=5.7$ on a $16^4$ lattice.
Red crosses, full(CLE) QCD; blue circles, phase-quenched QCD. Vertical magenta 
dotted lines are at $m_\pi/2$ and $m_N/3$.}
\label{fig:plaquette57}
}
\parbox{0.2in}{}
\parbox{2.75in}{
\epsfxsize=2.75in   
\epsffile{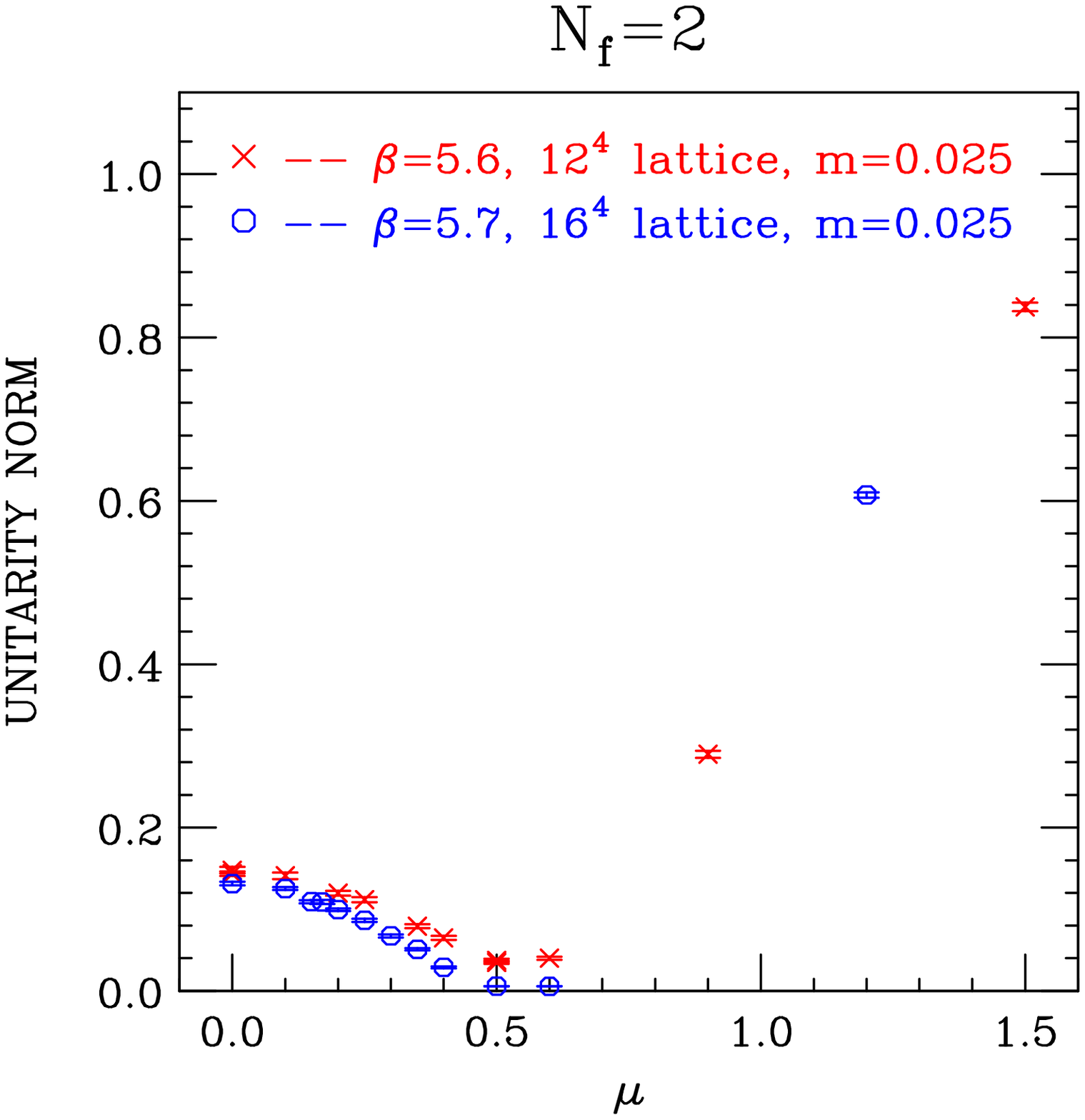}
\caption{Average unitarity norms as functions of $\mu$ for $\beta=5.6$,
$m=0.025$ on a $12^4$ lattice -- red. Same, but for $\beta=5.7$ on a $16^4$
lattice -- blue.}
\label{fig:unorm}
}
\end{figure}

In figure~\ref{fig:plaquette57} we plot the plaquette measured in these
simulations. The agreement between the 2 theories at $\mu=0$ is not as good as
for the other observables, however, it is significantly better than for
$\beta=5.6$. The plaquette stays close to its $\mu=0$ value through the
transition region. Again, the full (CLE) and phase-quenched results converge
and remain together for $\mu \ge 0.5$ whereas for $\beta=5.6$ a small difference
persists up to saturation.

In summary, we see some improvement in the behaviour of the CLE for $\beta=5.7$
over that for $\beta=5.6$. However, at this stage, we cannot rule out the
weak-coupling limit of the CLE producing phase-quenched results rather than
the correct results.

Finally in figure~\ref{fig:unorm} we show the average unitarity norm as a
function of $\mu$ from our simulations on a $16^4$ lattice at $\beta=5.7$ and
from our $12^4$ runs at $\beta=5.6$. We note that the unitarity norms for
$\beta=5.7$ are significantly smaller than those for  $\beta=5.6$. It has been
suggested that keeping unitarity norms small increases the chance that the CLE
will converge to the correct results, since there is less chance that the
trajectories will encounter zeros of the fermion determinant (poles in the
drift term). this suggests one should try even weaker couplings. Note also
that both these graphs indicate that the unitarity norm has a deep minimum
around $\mu=0.5$ or $0.6$. We suggest that the CLE simulations might converge
to the correct distributions for $\mu$ above or even slightly below these
minima, noting that this $\mu$ is approximately that for which the $\beta=5.7$
full and phase-quenched results converge.

\section{Summary and Discussion}

We simulate lattice QCD at finite quark-number chemical potential using the
complex Langevin equation (CLE) to determine whether this is a viable method
for evading the sign problems of this theory. Our simulations are performed at
$\beta=5.6$, $m=0.025$ on a $12^4$ lattice and at weaker coupling $\beta=5.7$,
$m=0.025$ on a $16^4$ lattice. We compare our observables with those obtained
in the phase-quenched approximation, where one uses only the magnitude of the
fermion determinant, using exact (RHMC) simulations. For $\mu < m_\pi/2$,
both theories should agree and the observables should remain fixed at their
$\mu=0$ values. For the full theory the observables should remain constant up
to $\mu \approx m_N/3$. Appreciably beyond $\mu=m_N/3$, the phase of the 
determinant should be small enough that both theories should be in agreement.

At $\mu=0$ the chiral condensate for $\beta=5.6$ shows sizable departures
from the known value, while at $\beta=5.7$ there is good agreement with the
exact result. In both cases, the condensate starts to fall (almost) immediately
$\mu > 0$, which does not agree with either the full or phase-quenched
expectations. However, the falloff for $\beta=5.7$ does appear to be slightly 
less rapid than at $\beta=5.6$. In the transition region the observables for
both $\beta$ values disagree both with what is expected and with those of the
phase-quenched approximation. At small $\mu$s through the transition region, 
the plaquettes for both $\beta=5.6$ and $\beta=5.7$ differ from the known
result for $\mu=0$ by a small but significant amount. However, the  $\beta=5.7$
plaquette is closer to the correct value than is the the $\beta=5.6$ value. 
For large $\mu$s the chiral condensates for both $\beta$s converge to their
phase-quenched values. In this regime the quark-number density and plaquette
for the $\beta=5.7$ simulations converge to the phase-quenched results while
for $\beta=5.6$ small but significant differences persist.

Hence, at $\beta=5.6$ and $\beta=5.7$ the CLE fails to produce correct results,
but some improvement is seen as we go to weaker coupling. However, the results
are worse than those of the phase-quenched approximation, so although it is
possible that in the weak-coupling limit, the CLE produces correct results, it
is also possible that it produces phase-quenched results. Therefore we are 
extending our simulations to even weaker couplings ($\beta=5.8$, $\beta=5.9$
and possibly $\beta=6.0$) on a $32^4$ lattice.

We also plan to research other (fermion) actions for which the CLE might 
exhibit better convergence. Because of flavour-symmetry (taste) breaking, the
zeros of the 2-flavour determinant have order $\frac{1}{2}$ instead of $2$.
This is likely to produce problems when calculating the chiral condensate or
quark-number density, both of which have poles at these zeros. (Is this why
the plaquettes appear to behave better over the transition region than the
chiral condensate and quark-number densities?) This suggests using Wilson
fermions whose exact vector flavour symmetry means that the 2-flavour
determinant will have zeros of order 2 as in the continuum. Another possible
action is one with an irrelevant chiral 4-fermion term, which moves the
(pseudo-)Goldstone pions to the auxiliary field and raises the mass of the
fermion fields, which produce the fermion determinant and have all the $\mu$
dependence \cite{Kogut:1998rg,Kogut:2006gt}. 
This can be thought of as replacing the current quarks with pions
and constituent quarks. Other possibilities include modifying the
gauge-fixing \cite{Nagata:2016alq}, 
or modifying the fermion dynamics by adding irrelevant terms to
the drift term \cite{Attanasio:2016mhc,Jaeger}. 
Both are designed to reduce the size of the unitarity norms.

Since the CLE appears to perform better in the nuclear matter (non-zero
quark-number density) phase, it might be possible to answer questions such as
if and under what circumstances such exotic phases as colour-superconductors
can be produced, even when the CLE fails in the region of the transition from
hadronic to nuclear matter.

We also plan to apply the CLE to simulate QCD at finite $\mu$ and temperature, 
$T$, since it has been observed that the CLE works better in this domain. Here
we will study the transition from hadronic and nuclear matter to a quark-gluon
plasma, and search for the critical endpoint.

\section*{Acknowledgments}

Our simulations are performed on Edison and Cori at NERSC under an ERCAP 
allocation, on Bridges at PSC, Comet at SDSC, and Stampede and Stampede~2
at TACC under an XSEDE allocation, on Blues and Bebop at LCRC, Argonne National
Laboratory and on Linux PCs at the HEP Division at Argonne National Laboratory.


\begin{thebibliography}{99}


\bibitem{Parisi:1984cs} 
  G.~Parisi,
  Phys.\ Lett.\ B {\bf 131}, 393 (1983).

\bibitem{Klauder:1983nn} 
  J.~R.~Klauder,
  Acta Phys.\ Austriaca Suppl.\  {\bf 25}, 251 (1983).

\bibitem{Klauder:1983zm} 
  J.~R.~Klauder,
  J.\ Phys.\ A {\bf 16}, L317 (1983).

\bibitem{Klauder:1983sp} 
  J.~R.~Klauder,
  Phys.\ Rev.\ A {\bf 29}, 2036 (1984).


\bibitem{Aarts:2009uq} 
  G.~Aarts, E.~Seiler and I.~O.~Stamatescu,
  Phys.\ Rev.\ D {\bf 81}, 054508 (2010)
  doi:10.1103/PhysRevD.81.054508
  [arXiv:0912.3360 [hep-lat]].

\bibitem{Aarts:2011ax} 
  G.~Aarts, F.~A.~James, E.~Seiler and I.~O.~Stamatescu,
  Eur.\ Phys.\ J.\ C {\bf 71}, 1756 (2011)
  doi:10.1140/epjc/s10052-011-1756-5
  [arXiv:1101.3270 [hep-lat]].

\bibitem{Nagata:2015uga} 
  K.~Nagata, J.~Nishimura and S.~Shimasaki,
  PTEP {\bf 2016}, no. 1, 013B01 (2016)
  doi:10.1093/ptep/ptv173
  [arXiv:1508.02377 [hep-lat]].

\bibitem{Nishimura:2015pba} 
  J.~Nishimura and S.~Shimasaki,
  Phys.\ Rev.\ D {\bf 92}, no. 1, 011501 (2015)
  doi:10.1103/PhysRevD.92.011501
  [arXiv:1504.08359 [hep-lat]].

\bibitem{Nagata:2016vkn} 
  K.~Nagata, J.~Nishimura and S.~Shimasaki,
  Phys.\ Rev.\ D {\bf 94}, no. 11, 114515 (2016)
  doi:10.1103/PhysRevD.94.114515
  [arXiv:1606.07627 [hep-lat]].

\bibitem{Aarts:2017vrv} 
  G.~Aarts, E.~Seiler, D.~Sexty and I.~O.~Stamatescu,
  JHEP {\bf 1705}, 044 (2017)
  doi:10.1007/JHEP05(2017)044
  [arXiv:1701.02322 [hep-lat]].

\bibitem{Seiler:2017wvd} 
  E.~Seiler,
  arXiv:1708.08254 [hep-lat].


\bibitem{Seiler:2012wz} 
  E.~Seiler, D.~Sexty and I.~O.~Stamatescu,
  Phys.\ Lett.\ B {\bf 723}, 213 (2013)
  [arXiv:1211.3709 [hep-lat]].


\bibitem{Aarts:2008rr} 
  G.~Aarts and I.~O.~Stamatescu,
  JHEP {\bf 0809}, 018 (2008)
  doi:10.1088/1126-6708/2008/09/018
  [arXiv:0807.1597 [hep-lat]].

\bibitem{Aarts:2013uxa} 
  G.~Aarts, L.~Bongiovanni, E.~Seiler, D.~Sexty and I.~O.~Stamatescu,
  Eur.\ Phys.\ J.\ A {\bf 49}, 89 (2013)
  doi:10.1140/epja/i2013-13089-4
  [arXiv:1303.6425 [hep-lat]].

\bibitem{Aarts:2014bwa} 
  G.~Aarts, E.~Seiler, D.~Sexty and I.~O.~Stamatescu,
  Phys.\ Rev.\ D {\bf 90}, no. 11, 114505 (2014)
  doi:10.1103/PhysRevD.90.114505
  [arXiv:1408.3770 [hep-lat]].

\bibitem{Aarts:2016qrv} 
  G.~Aarts, F.~Attanasio, B.~Jäger and D.~Sexty,
  JHEP {\bf 1609}, 087 (2016)
  doi:10.1007/JHEP09(2016)087
  [arXiv:1606.05561 [hep-lat]].

\bibitem{Langelage:2014vpa} 
  J.~Langelage, M.~Neuman and O.~Philipsen,
  JHEP {\bf 1409}, 131 (2014)
  doi:10.1007/JHEP09(2014)131
  [arXiv:1403.4162 [hep-lat]].



\bibitem{Sexty:2013ica} 
  D.~Sexty,
  Phys.\ Lett.\ B {\bf 729}, 108 (2014)
  [arXiv:1307.7748 [hep-lat]].

\bibitem{Fodor:2015doa} 
  Z.~Fodor, S.~D.~Katz, D.~Sexty and C.~Török,
  Phys.\ Rev.\ D {\bf 92}, no. 9, 094516 (2015)
  doi:10.1103/PhysRevD.92.094516
  [arXiv:1508.05260 [hep-lat]].

\bibitem{Nagata:2016mmh} 
  K.~Nagata, H.~Matsufuru, J.~Nishimura and S.~Shimasaki,
  PoS LATTICE {\bf 2016}, 067 (2016)
  [arXiv:1611.08077 [hep-lat]].


\bibitem{Sinclair:2016nbg} 
  D.~K.~Sinclair and J.~B.~Kogut,
  PoS LATTICE {\bf 2016}, 026 (2016)
  [arXiv:1611.02312 [hep-lat]].

\bibitem{Sinclair:2015kva} 
  D.~K.~Sinclair and J.~B.~Kogut,
  PoS LATTICE {\bf 2015}, 153 (2016)
  [arXiv:1510.06367 [hep-lat]].


\bibitem{Bloch:2016jwt} 
  J.~Bloch, J.~Glesaaen, O.~Philipsen, J.~Verbaarschot and S.~Zafeiropoulos,
  EPJ Web Conf.\  {\bf 137}, 07030 (2017)
  doi:10.1051/epjconf/201713707030
  [arXiv:1612.04621 [hep-lat]].

\bibitem{Nagata:2016alq} 
  K.~Nagata, J.~Nishimura and S.~Shimasaki,
  JHEP {\bf 1607}, 073 (2016)
  doi:10.1007/JHEP07(2016)073
  [arXiv:1604.07717 [hep-lat]].


\bibitem{Ukawa:1985hr}
  A.~Ukawa and M.~Fukugita,
  Phys.\ Rev.\ Lett.\  {\bf 55} (1985) 1854.

\bibitem{Fukugita:1986tg}
  M.~Fukugita, Y.~Oyanagi and A.~Ukawa,
  Phys.\ Rev.\ D {\bf 36} (1987) 824.

\bibitem{Fukugita:1988qs} 
  M.~Fukugita and A.~Ukawa,
  Phys.\ Rev.\ D {\bf 38}, 1971 (1988).


\bibitem{Brown:1991qw} 
  F.~R.~Brown, F.~P.~Butler, H.~Chen, N.~H.~Christ, Z.~h.~Dong, W.~Schaffer, 
  L.~I.~Unger and A.~Vaccarino,
  Phys.\ Rev.\ Lett.\  {\bf 67}, 1062 (1991).
  doi:10.1103/PhysRevLett.67.1062

\bibitem{Schaffer:1992rq} 
  W.~Schaffer,
  Nucl.\ Phys.\ Proc.\ Suppl.\  {\bf 30}, 405 (1993).
  doi:10.1016/0920-5632(93)90238-2



\bibitem{Kogut:2002zg} 
  J.~B.~Kogut and D.~K.~Sinclair,
  Phys.\ Rev.\ D {\bf 66}, 034505 (2002)
  doi:10.1103/PhysRevD.66.034505
  [hep-lat/0202028].


\bibitem{Sinclair:2006zm} 
  D.~K.~Sinclair and J.~B.~Kogut,
  PoS LAT {\bf 2006}, 147 (2006)
  [hep-lat/0609041].


\bibitem{Kogut:1998rg} 
  J.~B.~Kogut, J.~F.~Lagae and D.~K.~Sinclair,
  Phys.\ Rev.\ D {\bf 58}, 034504 (1998)
  doi:10.1103/PhysRevD.58.034504
  [hep-lat/9801019].

\bibitem{Kogut:2006gt} 
  J.~B.~Kogut and D.~K.~Sinclair,
  Phys.\ Rev.\ D {\bf 73}, 074512 (2006)
  doi:10.1103/PhysRevD.73.074512
  [hep-lat/0603021].


\bibitem{Attanasio:2016mhc} 
  F.~Attanasio and B.~Jäger,
  PoS LATTICE {\bf 2016}, 053 (2016)
  [arXiv:1610.09298 [hep-lat]].

\bibitem{Jaeger}
Benjamin J\"{a}ger, Felipe Attanasio and Gert Aarts,
{\it Improved convergence of Complex Langevin simulations}, in
{\it Proceedings, 35th International Symposium on Lattice Field Theory 
(Lattice2017): Granada, Spain,} to appear in EPJ Web Conf.


\end{thebibliography}
\end{document}